\documentclass[aip,graphicx,reprint,notitlepage]{revtex4-1}
\usepackage{amsmath}
\usepackage{amssymb}
\usepackage{graphicx}
\usepackage{svg}
\usepackage{hyperref}
\usepackage[caption=false]{subfig}
\usepackage{physics}

\draft 

\begin{document}

\title{Thermal transport of flexural phonons in a rectangular wire}

\author{G. Rivas Álvarez}
\email[]{rivas@ciencias.unam.mx}
\affiliation{Instituto de F\'isica, Universidad Nacional Aut\'onoma de M\'exico, 04510 Ciudad de M\'exico, Mexico}

\author{E. Benítez Rodríguez}
\affiliation{Departamento de F\'isica, Universidad Aut\'onoma Metropolitana-Iztapalapa, 09310 Iztapalapa, Ciudad de M\'exico, Mexico}

\author{M. A. Bastarrachea-Magnani}
\affiliation{Departamento de F\'isica, Universidad Aut\'onoma Metropolitana-Iztapalapa, 09310 Iztapalapa, Ciudad de M\'exico, Mexico}

\author{M. Martínez-Mares}
\affiliation{Departamento de F\'isica, Universidad Aut\'onoma Metropolitana-Iztapalapa, 09310 Iztapalapa, Ciudad de M\'exico, Mexico}

\author{R. A. M\'endez-S\'anchez}
\affiliation{Instituto de Ciencias F\'isicas, Universidad Nacional Aut\'onoma 
de M\'exico, 62210 Cuernavaca, Morelos, Mexico}

\begin{abstract}
We study the quantum thermal transport of elastic excitations through a two-dimensional elastic waveguide between two thermal reservoirs. We solve the classical Kirchhoff-Love equation for rectangular wires and explore the dispersion relation for both the symmetric and antisymmetric solutions. Then, we study the phonon transport of these modes within the second quantization framework by analyzing the mean quadratic displacement, from which the energy density current, the temperature field, and conductance are determined. We identify the relevant modes contributing to thermal transport and explore the average temperature difference to reach the high-temperature limit. We expect our results to pave the way for understanding phonon-mediated thermal transport in two-dimensional mesoscopic quantum devices. 
\end{abstract}

\maketitle 

\section{Introduction}
\label{sec:1}

Heat transport through mesoscopic quantum devices has remained an attractive 
topic for at least the last three decades. Much of the interest has relied on an 
expected analogy to the electrical 
conduction~\cite{Pendry,Maynard,Sivan,Butcher,Angelescu,Blencowe,Patton}. During 
those years, the quantization of thermal transport through ballistic 
constrictions at low temperatures has been predicted and observed 
~\cite{Rego1,Kaso}. The classical thermal conductance~\cite{Tighe,Yung,Chiatti} 
and its quantum correlate~\cite{Schwab,Meschke,Jezouin} have been measured, to 
which it was found a universal independence of particle 
statistics~\cite{Rego1,Rego2,Blencowe2,Krive}. Thermoelectric transport, the 
thermopower, and the Peltier coefficient, the thermal conductivity and its 
comparison with the electrical counterpart, that is, the Lorenz number, have 
been mesoscopic phenomena of interest for the investigation of quantum effects 
on physical quantities defined on the 
bulk~\cite{Sivan,Butcher,Chiatti,Esposito,Molenkamp,Greiner}. Recently, problems 
of thermal rectification, transistors, logical gates, and memory have also 
renewed the interest in mesoscopic heat transport at the nanoscale 
level~\cite{Chang,Li,Wang1,Wang2}.

With the advancement of technology, it has become possible to manufacture fully suspended three-dimensional nanostructured devices that can be used for thermal transport measurements. In this sense, a vast array of devices are employed, including single crystal one-dimensional wires~\cite{Tighe}, insulating nanostructures or membranes~\cite{Schwab}, normal metal films~\cite{Yung}, and superconducting leads~\cite{Meschke}. It is convenient that the operation depends on whether thermal conduction is activated by electrons~\cite{Chiatti}, photons~\cite{Meschke}, or phonons~\cite{Schwab}. Since it has been shown that the measured resonances of the steady-state resistance of small metal wires, influenced by an electric field, correspond to acoustic modes of the wire~\cite{Seyler}, a theoretical analysis of its acoustic modes becomes timely. Along this line, the spectrum of free rectangular thin plates was analyzed statistically, looking for fingerprints of wave chaos~\cite{LopezGlz}. The free vibration modes of a rectangular quantum wire composed of cubic crystals were obtained theoretically~\cite{Nishiguchi}. There, it has been found that four independent vibrational modes at zero temperature open four channels for heat transport. Similarly, using the thin-plate theory of elasticity, the complete mode spectrum of a rectangular beam was calculated in the long-wavelength limit~\cite{Cross}.

In the present work, we propose a study of phonon heat transport through a homogeneous rectangular quantum wire perfectly connected to two thermal reservoirs at different temperatures. This configuration is similar to that proposed in Ref.~\onlinecite{Kaso} for an inhomogeneous contact-device-contact system composed of an arbitrary oscillator chain but with the nonperfect coupling of the device to the contacts. In a similar configuration, the thin-plate theory of elasticity is used to treat the transmission of elastic waves between a cavity abruptly connected, through a bridge, to a thermal reservoir at a different temperature~\cite{Cross}. We start from the thin-plate theory to find the free-standing flexural modes for the displacement, a field that is quantized using the second quantization formalism. We are interested in the mean quadratic displacement, in the quantum statistical mechanics sense, that allows us to obtain the energy flux, the thermal conductance, and the temperature distribution along the wire. 

The article is organized as follows. In Sec.~\ref{sec:2}, we solve the classical Kirchhoff-Love equation for the rectangular plate and identify the symmetric and antisymmetric modes of propagating flexural vibration. Next, in Sec.~\ref{sec:3}, we quantize the classical solutions and study phonon thermal transport in the rectangular plate assuming a temperature gradient between two thermal reservoirs, and calculating the mean quadratic displacement, mean quadratic velocity, and energy current for each parity. Finally, in Sec.~\ref{sec:4}, we present our conclusions.

\section{Solution to the Kirchhoff-Love equation}
\label{sec:2}

\begin{figure}
     \begin{center}
     \subfloat[]{\includegraphics[width=1\columnwidth]{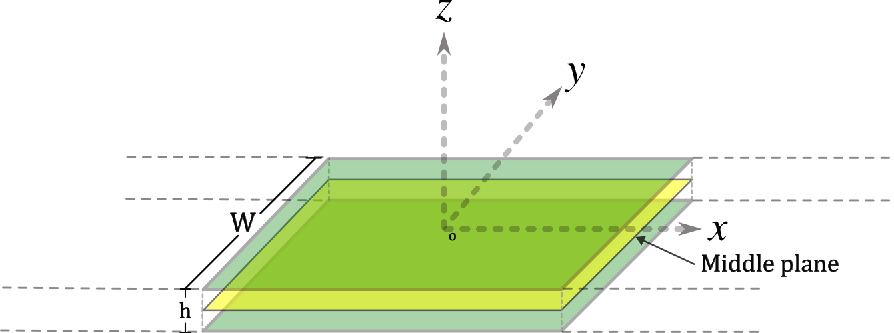}}\\
     \subfloat[]{\includegraphics[width=0.65\columnwidth]{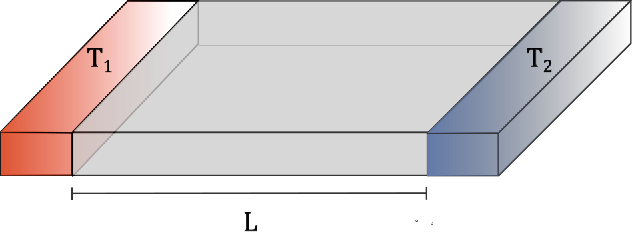}}
     \caption{(a) A thin plate of thickness $h$ on the $xy$ plane. The plate is finite in the $y$-direction with width $W$ and very long in the $x$-direction, where propagation occurs. (b) Schematic depiction of the thermal transport of phonons. Two reservoirs at different temperatures, $T_1$ and $T_2$, are placed and connected to a plate of length $L$.}
     \label{fig:plate}
     \end{center}
\end{figure}

\begin{figure}
    \centering
    \includegraphics[width=1.0\columnwidth]{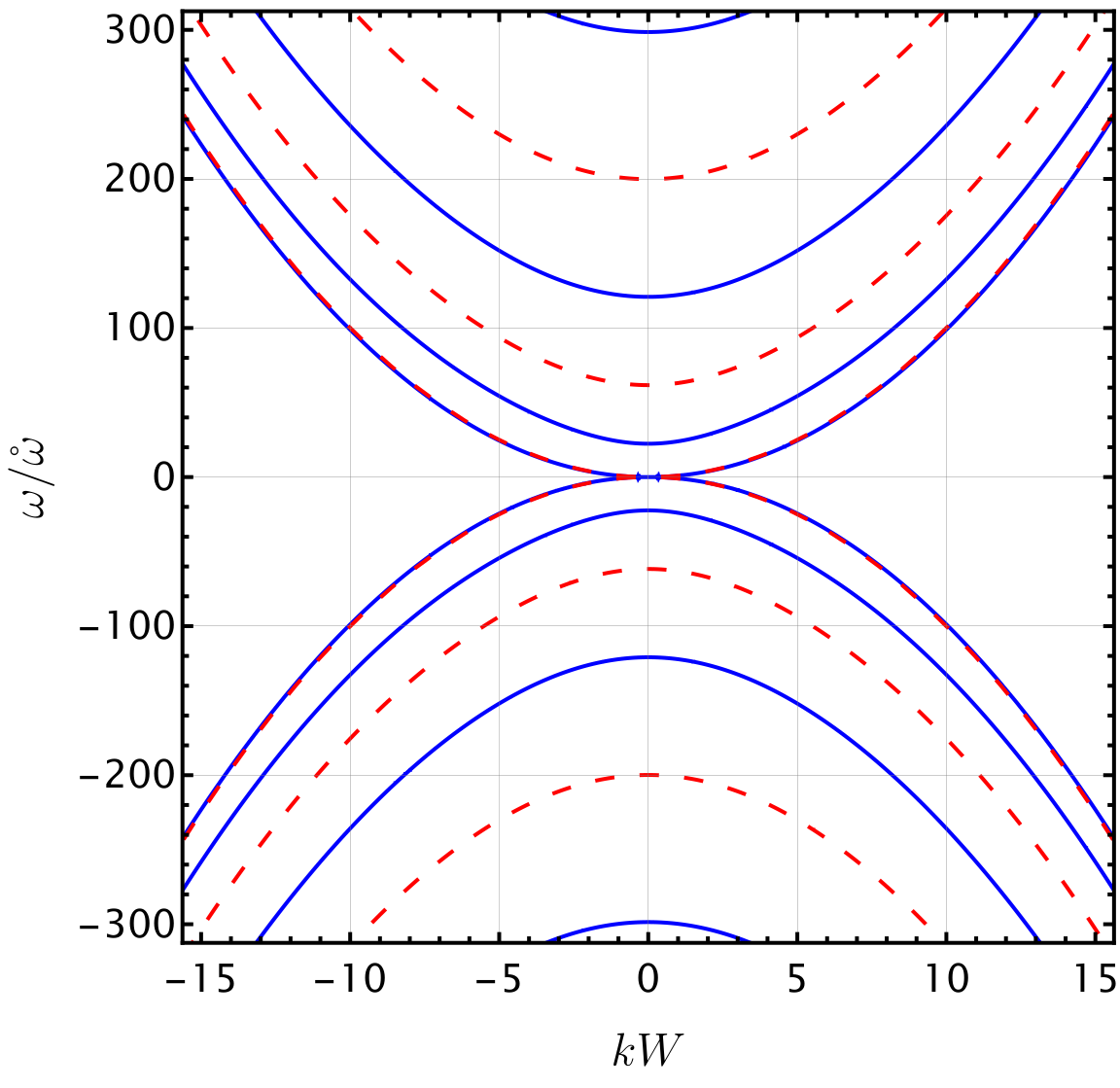}   
    \caption{Dispersion relations for the flexural modes in a rectangular wire with Poisson ratio $\nu = 1/3$, the mean value for aluminium and aluminium alloys~\cite{Chmelko2024}. The blue solid (red dashed) curves correspond to the symmetric (antisymmetric) solutions.}
    \label{fig:1}
\end{figure}

\begin{figure}
     \centering
     \subfloat[][Symmetric 
solution.]{\includegraphics[width=0.8\columnwidth]{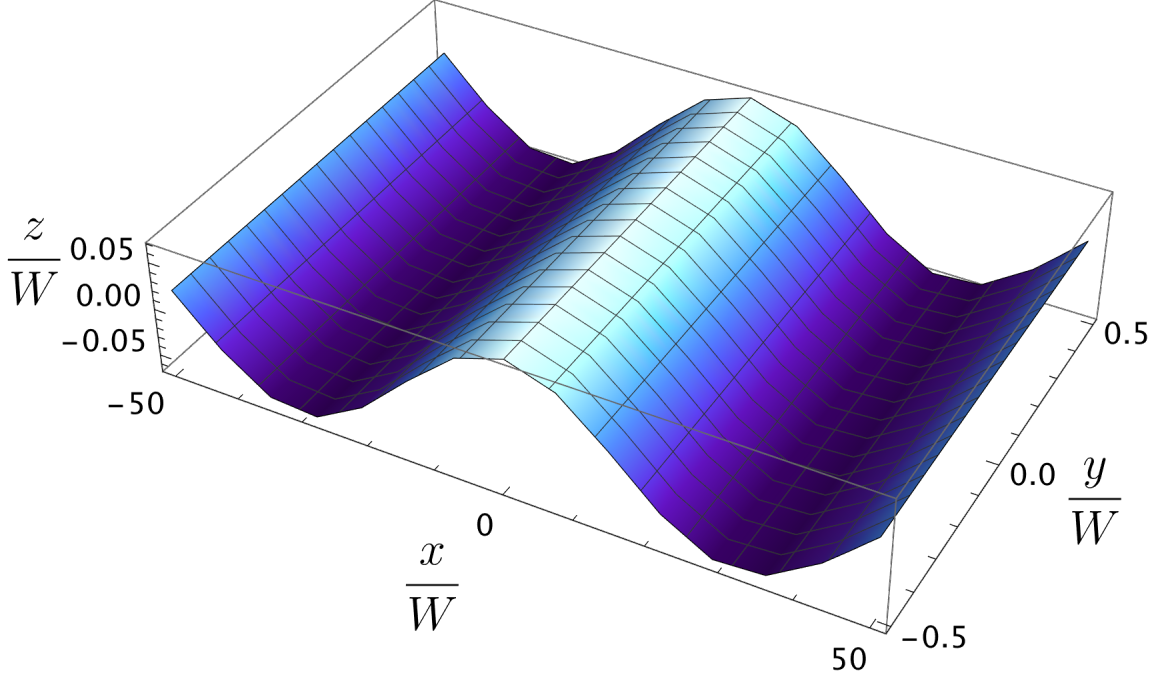}\label{fig:}}\\
     \subfloat[][Antisymmetric 
solution.]{\includegraphics[width=0.8\columnwidth]{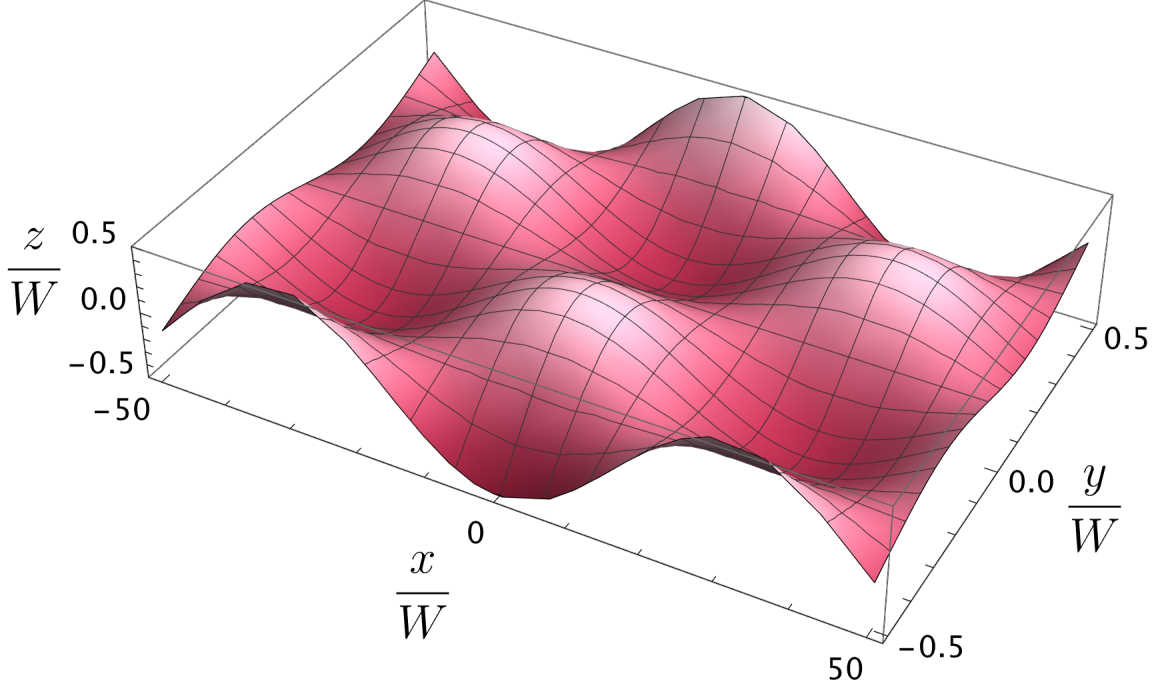}}\label{<figure2>}
     \caption{Example of the first (a) symmetric and second (b) antisymmetric flexural wave mode for $kW = 0.1$.}
     \label{fig:steady_state}
\end{figure}

\begin{figure}[h]
    \centering
    \includegraphics[width=1.0\columnwidth]{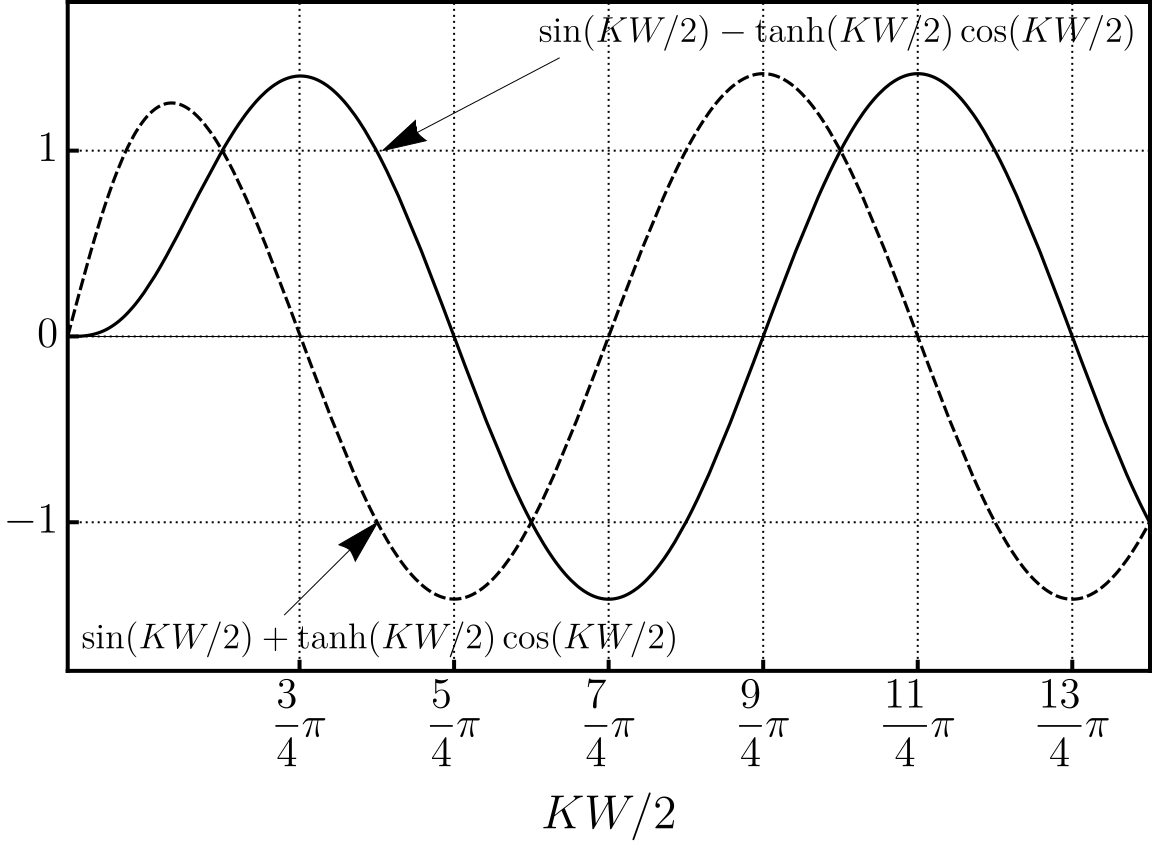}
    \caption{Plot of the function $\sin\left(K_{m}W/2 \right) \pm \tanh\left(K_{m}W/2\right) \cos\left(K_{m}W/2\right)$ that is employed to find the cut-off frequencies $\omega_{m}$ as roots of Eq.~\ref{eq:zeroWaveN}.}
    \label{fig:zeroWaveN_Plot}
\end{figure}

Consider a thin plate with thickness $h$ on the $yz$ plane, width $W$ along the 
$y$ axis, and extending through all the $x$-axis, as illustrated in 
Fig.~\ref{fig:plate}. Assuming that plane waves traveling along the $x$-axis 
have a wavelength much larger than $h$, the low-frequency equation for the 
vertical displacement $u_{z}\left(x,y;t\right)$, in the absence of external 
forces, is the Kirchhoff-Love equation~\cite{Levan2017,Ozenda2021,Nguyen2021},
\begin{equation}\label{eq:Kirchhoff}
    \left(\nabla_{\perp}^{4}+\frac{\rho h}{D}\frac{\partial^{2}}{\partial t^{2}}\right)u_{z}\left(x,y;t\right)=0,
\end{equation}
where $\rho$ is the density of the plate; $D=Eh^{3}/12(1-\nu^{2})$, it is the flexural rigidity, with $E$ and $\nu$ being the Young's modulus and Poisson's ratio, respectively; and
$\nabla_{\perp}^{4}=\nabla_{\perp}^{2}\nabla_{\perp}^{2}$, with 
$\nabla_{\perp}^{2}$ the two-dimensional Laplacian operator. For free edges at 
$y=\pm W/2$, the boundary conditions are
\begin{subequations}
\label{eq:Boundary}
\begin{align}
    \left.\left(\frac{\partial^{2}u_{z}}{\partial y^{2}}+\nu\frac{\partial^{2}u_{z}}{\partial x^{2}}\right)\right|_{y=\pm W/2} & =0,
    \label{eq:Boundary1} \\
    \left.\left[\frac{\partial^{3}u_{z}}{\partial y^{3}}+\left(2-\nu\right)\frac{\partial^{3}u_{z}}{\partial y\partial x^{2}}\right]\right|_{y=\pm W/2} & =0.
    \label{eq:Boundary2}
\end{align}
\end{subequations}
For an infinite waveguide, the Kirchhoff-Love equation, 
Eq.~\eqref{eq:Kirchhoff}, is separable; therefore, an ansatz for flexural waves 
traveling along the $x$-axis is 
$u_{z}(x,y;t)=w(y)\mathrm{e}^{\mathrm{i}(kx-\omega t)}$, where $k$ and $\omega$ 
are the wave number and frequency, respectively. The amplitude $w(y)$ represents 
a scalar field giving the shape of the flexural vibration in the $y$-direction 
satisfying
\begin{equation}
    \left[\dv[4]{y}-2k^{2}\dv[2]{y}+\left(k^{4}-K^{4}\right)\right]w\left(y\right)=0,\label{eq:LoveEquationForW}
\end{equation}
where the dispersion relation is given by
\begin{equation}
\omega(k)=\pm\sqrt{D/h\rho}K^{2}(k).
\label{eq:disp.rel}
\end{equation}
The boundary conditions, Eqs.~\eqref{eq:Boundary1} and~\eqref{eq:Boundary2}, 
are written as
\begin{subequations}
\label{eq:BConditionW}
\begin{align}
    \left.\left[\dv[2]{w(y)}{y}-\nu k^{2}w(y)\right]\right|_{y=\pm W/2} & =0,
    \label{eq:BConditionW1}\\
    \left.\left[\dv[3]{w(y)}{y}-\left(2-\nu\right)k^{2}\dv{w(y)}{y}\right]\right|_{y=\pm W/2} & =0.
    \label{eq:BConditionW2}
\end{align}
\end{subequations}
Let $w(y)=\mathrm{e}^{\mathrm{i}\lambda y}$ an ansatz to the fourth order 
differential equation, Eq.~\eqref{eq:LoveEquationForW}. This leads to a 
fourth-grade algebraic equation for $\lambda$,
\begin{equation}
    \lambda^{4}+2k^{2}\lambda^{2}+\left(k^{4}-K^{4}\right)=0,
    \label{eq:algebraicEqW}
\end{equation}
with solutions $\lambda=\pm\sqrt{K^{2}-k^{2}}$, 
$\pm\mathrm{i}\sqrt{K^{2}+k^{2}}$, which allow to construct general solutions to 
Eq~\eqref{eq:LoveEquationForW} in terms or trigonometric (hyperbolic) functions. 
By considering the definite parity of $w(y)$, the symmetric and antisymmetric 
solutions of Eq.~\eqref{eq:LoveEquationForW} are
\begin{eqnarray}
    w^{(s)}(y) & = & a\cosh\left(\sqrt{K^{2}+k^{2}}y\right) 
    \nonumber \\ & + & 
    b\cos\left(\sqrt{K^{2}-k^{2}}y\right),
    \label{eq:wSymmetricSol}  \\
    w^{(a)}(y) & = & a'\sinh\left(\sqrt{K^{2}+k^{2}}y\right) 
    \nonumber \\ & + & 
    b'\sin\left(\sqrt{K^{2}-k^{2}}y\right),
    \label{eq:wAntiSymmetricSol}
\end{eqnarray}
where $a$, $b$, $a'$, $b'$ are set by the boundary conditions, 
Eqs.~\eqref{eq:BConditionW1} and~\eqref{eq:BConditionW2}. The symmetric and 
antisymmetric solutions lead to
\begin{widetext}
    \begin{eqnarray}
        \left[ \begin{array}{cc}
        \kappa_{+}\cosh\left(\sqrt{K^{2}+k^{2}}\frac{W}{2}\right) & -\kappa_{-}\cos\left(\sqrt{K^{2}-k^{2}}\frac{W}{2}\right) \\
        \kappa_{-}\sqrt{K^{2}+k^{2}}\sinh\left(\sqrt{K^{2}+k^{2}}\frac{W}{2}\right) & \kappa_{+}\sqrt{K^{2}-k^{2}}\sin\left(\sqrt{K^{2}-k^{2}}\frac{W}{2}\right)
        \end{array} \right] 
        \left( \begin{array}{c}
        a \\ b
        \end{array} \right) & = &
        \left(\begin{array}{c}
        0 \\ 0
        \end{array}
        \right),
        \label{eq:SymmetricBC} \\
        \left[\begin{array}{cc}
        \kappa_{+}\sinh\left(\sqrt{K^{2}+k^{2}}\frac{W}{2}\right) & -\kappa_{-}\sin\left(\sqrt{K^{2}-k^{2}}\frac{W}{2}\right)\\
        \kappa_{-}\sqrt{K^{2}+k^{2}}\cosh\left(\sqrt{K^{2}+k^{2}}\frac{W}{2}\right) & -\kappa_{+}\sqrt{K^{2}-k^{2}}\cos\left(\sqrt{K^{2}-k^{2}}\frac{W}{2}\right)
        \end{array} \right] 
        \left(\begin{array}{c}
        a' \\ b'
        \end{array} \right) & = & 
        \left(\begin{array}{c}
        0 \\ 0
        \end{array}\right), 
        \label{eq:ASymmetricBC}
    \end{eqnarray}
respectively, where $\kappa_{\pm}=K^{2}\pm(1-\nu)k^{2}$. To avoid trivial 
solutions to the boundary conditions, Eqs.~\eqref{eq:SymmetricBC} 
and~\eqref{eq:ASymmetricBC}, their determinant must vanish. This condition leads 
to 
\begin{eqnarray}
       & & \kappa_{+}^{2}\sqrt{K^{2}-k^{2}}\cosh\left(\sqrt{K^{2}+k^{2}}\frac{W}{2}\right)\sin\left(\sqrt{K^{2}-k^{2}}\frac{W}{2}\right) 
        \nonumber \\ & + & 
        \kappa_{-}^{2}\sqrt{K^{2}+k^{2}}\sinh\left(\sqrt{K^{2}+k^{2}}\frac{W}{2}\right)\cos\left(\sqrt{K^{2}-k^{2}}\frac{W}{2}\right) = 0,
        \label{eq:DispersionRelationS} 
        \\ & &
        \kappa_{-}^{2}\sqrt{K^{2}+k^{2}}\cosh\left(\sqrt{K^{2}+k^{2}}\frac{W}{2}\right)\sin\left(\sqrt{K^{2}-k^{2}}\frac{W}{2}\right) 
        \nonumber \\ & - & 
        \kappa_{+}^{2}\sqrt{K^{2}-k^{2}}\sinh\left(\sqrt{K^{2}+k^{2}}\frac{W}{2}\right)\cos\left(\sqrt{K^{2}-k^{2}}\frac{W}{2}\right) = 0,
        \label{eq:DispersionRelationA}
    \end{eqnarray}
\end{widetext}
for the symmetric and antisymmetric solutions, respectively. Since $K$ is a 
function of the frequency $\omega$, Eqs.~\eqref{eq:DispersionRelationS} 
and~\eqref{eq:DispersionRelationA} must be interpreted as dispersion relations. 
The dispersion relations are shown in Fig.~\ref{fig:1}, where we take 
$\mathring{\omega}=W^{-2}\sqrt{D/\rho h}$ as a frequency scale.
Finally, the full solutions become
\begin{gather}
u^{(s)}(x,y;t) = \sum_{m} w_{m}^{(s)} \mathrm{e}^{\mathrm{i}(k_{m}x-\omega_{m}t)}, \\
u^{(a)}(x,y;t) = \sum_{m} w_{m}^{(a)} \mathrm{e}^{\mathrm{i}(k_{m}x-\omega_{m}t)}.
\end{gather}
The first symmetric and antisymmetric modes for $kW=0.1$ are shown in 
Fig.~\ref{fig:steady_state}.
The vanishing wavenumbers, $k_{m}(\omega_{m})=0$, are of particular interest: 
they define the cut-off angular frequencies $\omega_m$ at which the flexural 
modes start opening. Setting $k=0$ in the dispersion relations, 
Eqs.~\eqref{eq:DispersionRelationS} and~\eqref{eq:DispersionRelationA}, they 
reduce to
\begin{equation}
    \sin\left(K_{m}W/2 \right) \pm \tanh\left(K_{m}W/2 \right) \cos\left(K_{m}W/2 \right)=0.
    \label{eq:zeroWaveN}
\end{equation}
Both functions on the left-hand side of this equation are plotted in 
Fig.~\ref{fig:zeroWaveN_Plot} so that the roots are at the crossing with the 
$x$-axis. It can be seen that the first root occurs at $K_1=0$, which gives 
$\omega_1=0$. The non-zero roots of Eq.~\eqref{eq:zeroWaveN} are very close to
\begin{equation}
    \frac{K_{m-1}W}{2} = \frac{\left(2m-1 \right)\pi}{4}, 
    \label{eq:zeroWaveN_Kn}
\end{equation}
for $m>2$, since $\tanh(K_{m}W/2)$ tends quickly to unit. In terms of the 
angular frequency,
\begin{equation}
    \omega_{m-1}=\frac{\left(2m-1 \right)^{2}\pi^{2}}{4W^{2}}\sqrt{\frac{D}{\rho h}}.
    \label{eq:zeroWaveN_omega}
\end{equation}

\section{Thermal phonon transport}
\label{sec:3}

Next, we consider the thin plate to be located between two reservoirs with 
different temperatures separated by a distance $L$ [see 
Fig.~\ref{fig:plate}(b)]. As a result, the system enters a non-equilibrium state 
that we assume is stationary. To find it, we propose a simple model for the 
reduced non-equilibrium stationary state density matrix of quantized acoustic 
excitations that transport energy between the two reservoirs. First, we write 
the vertical displacement as
\begin{equation}
u_{z}(x,y,t) = a_{k}(t) \Phi_{k}(x,y),
\end{equation}
and impose periodic boundary conditions in the $x-$direction, so that 
$k=k_{\ell}=\frac{2\pi \ell}{L}$, where $\ell=\pm 1,\pm 2,\dots,\pm N$, with $N$ 
giving the cut-off frequency. So, the most general displacement field is
\begin{equation}
\xi(x,y,t) = \sum_{\ell} \left[
a_{k_{\ell}}(t)\Phi_{k_{\ell}}(x,y)+a_{k_{\ell}}^{*}(t)\Phi_{k_{\ell}}^{*}(x,y) \right].
\label{eq:displacementField}
\end{equation}
The next step is to quantize the displacement field~\cite{Landau1980}. We 
associate 
$a_{k_{\ell}}(t)\to\sqrt{\hbar/2\rho\omega(k_{\ell})}\hat{a}_{k_{\ell}}$, where 
$\hat{a}_{k_{\ell}}$ ($\hat{a}_{k_{\ell}}^{\dagger}$) are annihilation 
(creation) phonon operators over the normalized states 
$\Phi_{k_{\ell}}(x,y)=C_{k_{\ell}}w_{k_{\ell}}(y)\mathrm{e}^{-\mathrm{i}k_{\ell}
x}$, $\int_{-h/2}^{h/2}\int_{-W/2}^{W/2}\int_{-L/2}^{L/2} \left| 
\Phi_{k_{\ell}}(x,y)\right|^{2}\dd x\,\dd y\,\dd z = 1$,  that describe the 
spatial profiles of the acoustic modes obtained in the previous section, and 
$w_{k_{\ell}}(y)$, playing the role of the polarization component in direction 
$y$ (flexural modes vibrate transversely to the propagation occurring in the 
$x-$direction). Hence, we turn the displacement field into an operator
\begin{equation}
\hat{\xi}(x,y) = \sum_{k_{l}}\sqrt{\frac{\hbar}{2\rho\,\omega(k_{l})}}\left[\hat{a}_{k_{l}}\Phi_{k_{l}}(x,y) +\hat{a}_{k_{l}}^{\dagger}\Phi_{k_{l}}^{*}(x,y)\right].
\label{eq:amplitudeOperator}
\end{equation}
Here, the sum is over the momentum in the propagation axis of the modes, so 
$\mathbf{k}=(k_{\ell},0,0)$, and the frequency in 
Eq.~\eqref{eq:amplitudeOperator} obeys the dispersion relations found earlier in 
Eq.~\eqref{eq:disp.rel}, and a Hamiltonian in canonical form
\begin{equation}
\hat{H}_{\text{ph}}=\sum_{k_{\ell}}\hbar\omega(k_{\ell})\left(\hat{a}_{k_{\ell}}^{\dagger}\hat{a}_{k_{\ell}}+\frac{1}{2}\right).
\end{equation}
describing a set of independent phonons in the second quantization 
formalism~\cite{Fetter1971}.

Next, we consider that the thin plate is in contact with two thermal reservoirs 
with different temperatures $k_{B}T_{i}=\beta_{i}^{-1}$, $i=1,2$, with $k_B$ the 
Boltzmann constant, as illustrated in Fig.~\ref{fig:plate} (b). Hence, we 
separate the Hamiltonian into those phonons traveling from $T_{1}$ to $T_{2}$ 
($k_{\ell}>0$) and vice versa ($k_{\ell}<0$), and consider a separable thermal 
state for each direction, so the density matrix of the thermal system is given 
by
\begin{eqnarray}
    \rho & = & \frac{1}{\mathcal{Z}_{+}}n_{\beta_{1}}
    \left[\sum_{k_{\ell}>0}\hbar\omega(k_{\ell})\hat{a}_{k_{\ell}}^{\dagger}\hat{a}_{k_{\ell}}\right]
    \nonumber \\ & \times &
    \frac{1}{\mathcal{Z}_{-}}n_{\beta_{2}}
    \left[\sum_{k_{\ell}<0}\hbar\omega(k_{\ell})\hat{a}_{k_{\ell}}^{\dagger}\hat{a}_{k_{\ell}}\right],
    \label{eq:densityMatrix}
\end{eqnarray}
where $n_{\beta_{i}}(x)=[\exp(\beta_{i}x)-1]^{-1}$ is the Bose-Einstein 
distribution for temperature $T_{i}$, and $\mathcal{Z}_{\pm}$ are the partition 
functions. In the case of large temperature to the frequency mode 
[$\hbar\omega(k_{\ell})\ll k_{B}T_{i}$], we can approximate it to its classical 
limit
\begin{eqnarray}
    \rho & = & \frac{1}{\mathcal{Z}_{+}} \exp\left(-\beta_{1}\sum_{k_{\ell}>0}\hbar\omega_{k_{\ell}}\hat{a}_{k_{\ell}}^{\dagger}\hat{a}_{k_{\ell}}\right) 
    \nonumber \\ & \times & 
    \frac{1}{\mathcal{Z}_{-}}
    \exp\left(-\beta_{2}\sum_{k_{\ell}<0}\hbar\omega_{k_{\ell}}\hat{a}_{k_{\ell}}^{\dagger}
    \hat{a}_{k_{\ell}}\right).
    \label{eq:densityMatrixHighT}
\end{eqnarray}
The density matrix allows us to find the thermal average value of any physical observable $\mathcal{O}$, associated with the operator $\hat{\mathcal{O}}$, through its mean value in the quantum statistical mechanics sense, that is, $\langle\hat{\mathcal{O}}\rangle=\mathrm{Tr}\left(\rho \mathcal{O}\right)$. Of particular interest are the mean quadratic displacement, $\langle\hat{\xi}^{2}\rangle$; the mean quadratic velocity, $\langle\hat{\dot{\xi}}^{2}\rangle$, associated with the plate's temperature profile; and the energy flux, $\langle \hat{J}\rangle$. 

For the mean quadratic displacement, we have 
\begin{widetext}
\begin{multline}
    \langle \hat{\xi}^{2}\rangle = \frac{\hbar}{2\rho} \sum_{\ell,\ell^{\prime}} 
    \sqrt{\frac{1}{\omega\left( k_{\ell} \right) \omega\left( k_{\ell^{\prime}} \right)}} \left[\left\langle\hat{a}_{k_{\ell}}^{\dagger}\hat{a}_{k_{\ell^{\prime}}}^{\dagger}\right\rangle\Phi_{k_\ell}^{*}\Phi_{k_{\ell^{\prime}}}^{*} + 
    \left\langle\hat{a}_{k_{\ell}}^{\dagger}\hat{a}_{k_{\ell^{\prime}}}\right\rangle\Phi_{k_\ell}^{*}\Phi_{k_{\ell^{\prime}}} + \left\langle\hat{a}_{k_{\ell}}\hat{a}_{k_{\ell^{\prime}}}^{\dagger}\right\rangle\Phi_{k_{\ell}}\Phi_{k_{\ell^{\prime}}}^{*} + 
    \left\langle\hat{a}_{k_{\ell}}\hat{a}_{k_{\ell^{\prime}}}\right\rangle\Phi_{k_{\ell}}\Phi_{k_{\ell^{\prime}}} \right].
    \label{eq:mqd1}
\end{multline}
\end{widetext}
Carrying out the trace over Fock states we find 
$\langle\hat{a}_{k_{\ell}}^{\dagger}\hat{a}_{k_{\ell^{\prime}}}^{\dagger}\rangle
=0=\langle\hat{a}_{k_{\ell}}\hat{a}_{k_{\ell^{\prime}}}\rangle$, and 
$\langle\hat{a}_{k_{\ell}}^{\dagger}\hat{a}_{k_{\ell^{\prime}}}\rangle=n_{\beta_
{i}}\left(\omega\left( k_{\ell}\right) \right) \delta_{\ell,\ell^{\prime}}$, 
where $\delta_{\ell,\ell^{\prime}}$ is the Kronecker's delta, and 
$n_{\beta_{i}}\left(\omega\left( k_{\ell}\right)\right)$ is the Bose-Einstein 
distribution
\begin{equation}
    n_{\beta_{i}}\left( \omega \left( k_{\ell} \right) \right)= 
    \frac{1}{\mathrm{e}^{\hbar\omega(k_{\ell})/k_{B}T_{i}}-1}.
\end{equation}
The temperature $T_{i}$ in the statistical distribution depends on $\mathrm{sgn(\ell)}$, i.e., the direction of the $\ell$-th mode: it is $T_{1}$ for right-moving phonons and  $T_{2}$ for the left-moving ones. With this and the commutator  
$[\hat{a}_{k_{\ell}},\hat{a}_{k_{\ell'}}^{\dagger}]=\delta_{\ell\ell^{\prime}}\
mathbb{I}$, the mean quadratic displacement, Eq.~\eqref{eq:mqd1}, is reduced to
\begin{eqnarray}
     \left\langle \hat{\xi}^{2}(y)\right\rangle & = & \frac{\hbar}{\rho} \left\{\sum_{\ell=1}^{N}
     \frac{C_{k_{\ell}}^{2}w_{k_{\ell}}^{2}(y)}{\abs{\omega( k_{\ell})}} 
     \left[ n_{\beta_{1}}(\omega(k_{\ell})) + 1 \right] \right.
     \nonumber \\ & + &
     \left. \sum_{\ell=-N}^{-1}
     \frac{C_{k_{\ell}}^{2}w_{k_{\ell}}^{2}(y)}{\abs{\omega( k_{\ell})}} 
     \left[ n_{\beta_{2}}(\omega(k_{\ell})) + 1 \right]\right\}.
     \label{eq:mqd2}
\end{eqnarray}
Notice that the mean quadratic displacement no longer depends on the $x$ coordinate. Invoking the symmetry of both the angular frequency $\omega(k_{\ell})$ and the amplitude $w_{\ell}(y)$ with respect to sign of the index $\ell$, the mean quadratic displacement is rewritten as,
\begin{equation}
    \left\langle \hat{\xi}^{2}(y) \right\rangle = \frac{\hbar}{\rho} \sum_{\ell=1}^{N}
     \frac{C_{k_{\ell}}^{2}w_{k_{\ell}}^{2}(y)}{\abs{\omega( k_{\ell})}} 
     [ n_{\beta_{1}}(\omega(k_{\ell})) + n_{\beta_{2}}(\omega(k_{\ell}))+2 ]
     \label{eq:meanQuadraticDisplacement}
\end{equation}
The temperature profile is defined with the mean quadratic velocity, $\langle \hat{\dot{\xi}}^{2}\rangle = \mathrm{Tr}(\rho\hat{\dot{\xi}}^{2})$, where the velocity operator is given by,
\begin{equation}
    \dot{\xi}(x,y) = \sum_{\ell=1}^{N} 
    \sqrt{\frac{\hbar \omega\left( k_{\ell}\right)}{2\rho}} \mathrm{i}
    \left[ \hat{a}_{k_{\ell}}^{\dagger}\Phi_{k_{\ell}}^{*}\left(x,y\right) - \hat{a}_{k_{\ell}}\Phi_{k_{\ell}}\left(x,y\right)\right].
    \label{eq:velocityOp}
\end{equation}
Following an analogous procedure as with the mean quadratic displacement, for the mean quadratic velocity, we find, 
\begin{eqnarray}
     \left\langle \hat{\dot{\xi}}^{2} \left( y\right)\right\rangle & = & \frac{\hbar}{\rho} 
     \sum_{\ell=1}^{N}\abs{\omega\left( k_{\ell}\right)}C_{k_{\ell}}^{2}w_{k_{\ell}}^{2}(y)
     \nonumber \\ & & \times 
     \left[n_{\beta_{1}}\left( \omega(k_{\ell})\right)+n_{\beta_{2}}\left( \omega(k_{\ell})\right)+ 2 \right].
     \label{eq:mqv}
\end{eqnarray}
Finally, the thermodynamic mean value of the energy current, $\langle \hat{J}\rangle$, is found with the energy current operator, 
\begin{equation}
    \hat{J}\left( x, y \right) =  -\frac{\rho c^{2}}{2} \left( \hat{\dot{\xi}} \dv{\hat{\xi}}{x} + \dv{\hat{\xi}}{x} \hat{\dot{\xi}}\right), 
    \label{eq:currentOperator}
\end{equation}
where $c$ is the velocity of the flexural mode. Following the same steps as with the mean quadratic displacement and the mean quadratic velocity, we find
\begin{equation}
    \langle \hat{J}(y)\rangle = \hbar c^{2} \sum_{\ell=1}^{N} k_{\ell} C_{k_{\ell}}^{2}w_{k_{\ell}}^{2}(y)\left[ n_{\beta_{1}}(\omega(k_\ell)) - n_{\beta_{2}}(\omega(k_\ell)) \right].
    \label{eq:energyCurrentDensity}
\end{equation}

The energy current is obtained by integrating the energy density current, Eq. 
\ref{eq:energyCurrentDensity}, over the cross-section, so that the  thermal 
conductance is defined by
\begin{equation}
    \mathcal{C} = \frac{h}{\Delta T}
    \int_{-W/2}^{W/2} \langle \hat{J}\left( y\right) \rangle \,\dd y,
    \label{eq:conductanceDefinition}
\end{equation}
where $\Delta T$ is the temperature difference between the reservoirs. Substituting the energy density current expression, Eq. ~\ref{eq:energyCurrentDensity}, 
\begin{equation}
    \mathcal{C} = \frac{\hbar c^{2}}{L\Delta T} \sum_{\ell=1}^{N} k_{\ell} 
    \left[ n_{\beta_{1}}\left( \omega(k_\ell)\right) - n_{\beta_{2}}\left( \omega(k_\ell)\right) \right],
\end{equation}
where we have taken into account that
\begin{equation}
    \int_{-W/2}^{W/2}C^{2}_{k_{\ell}}w_{k_{\ell}}^{2}(y)\, \dd y=\frac{1}{hL},
\end{equation}
because the normalization condition for $\Phi_{k_{\ell}}(x,y)$.

To study the conductance as a function of temperature, assume a small 
temperature difference $\Delta T \ll 1$K. This assumption simplifies the 
difference between statistical distributions, obtaining,
\begin{equation}
    \mathcal{C} = \frac{\hbar^{2} c^{2}}{Lk_{B}T^{2}} \sum_{\ell=1}^{N} k_{\ell} \omega(k_{\ell})
    \frac{\mathrm{e}^{\hbar\omega/k_{B}T}}
    {\left(\mathrm{e}^{\hbar\omega/k_{B}T}-1\right)^{2}},
\end{equation}
where $T=(T_{1}+T_{2})/2$ is the average temperature between the reservoirs. It is interesting to note that the thermal conductance depends on the inverse of the length of the wire, just as the electrical conductance of a one-dimensional resistor\cite{Gao}, except that this ``ohmic'' behavior seems to hold at any temperature range. In general, the thermal conductance changes with temperature. But, for a very large temperature ($\hbar\omega/k_BT\ll 1$), the thermal conductance becomes independent of the temperature. On the contrary, in the limit of very low temperature, $\hbar\omega/k_BT\gg 1$, the thermal conductance behaves as $\mathcal{C}\sim\exp(-\hbar\omega/k_BT)/T^2$. 

\begin{figure*}
    \centering
    \includegraphics[width=0.8\textwidth]{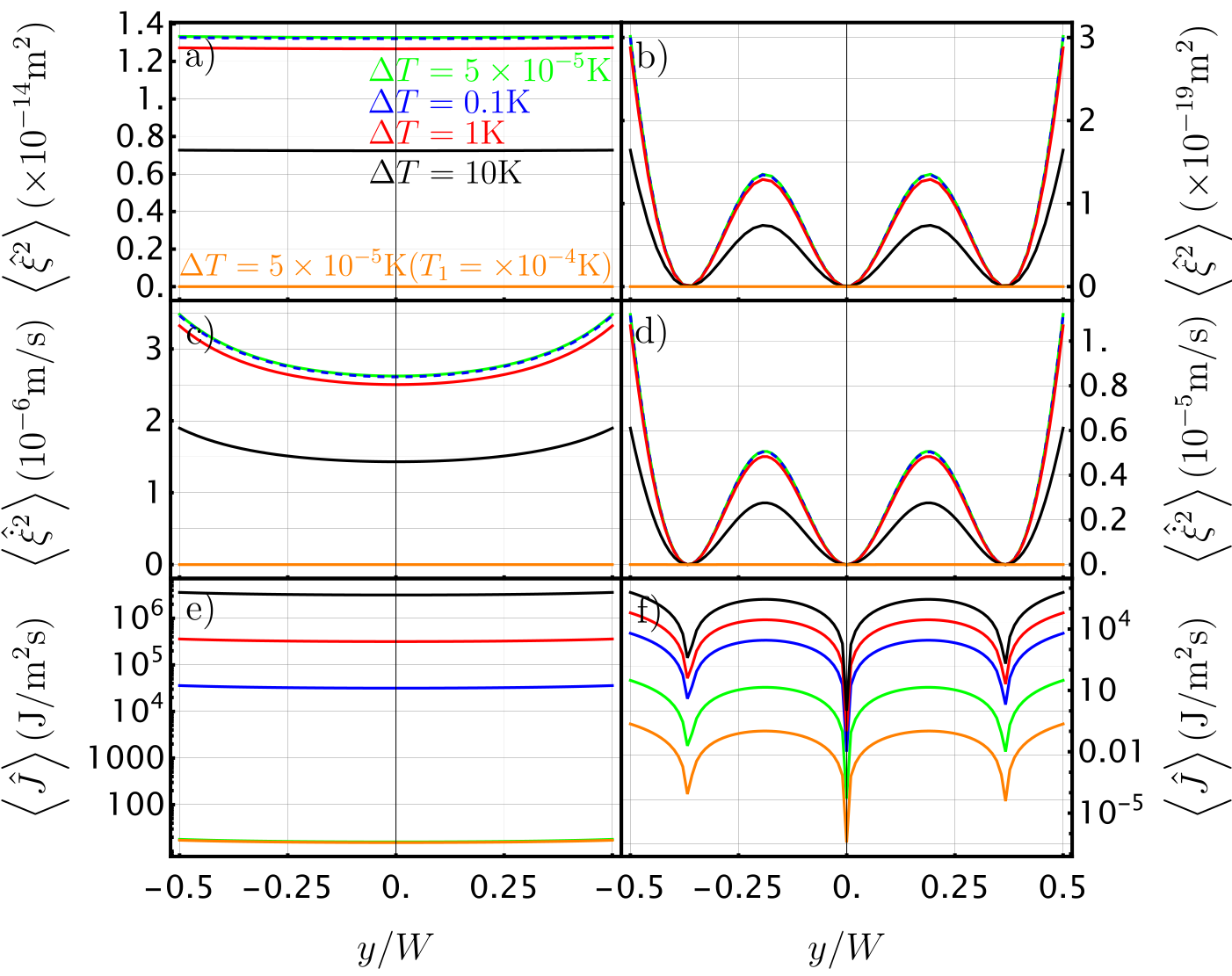}
    \caption{Transversal profiles of the different observables for the symmetric (left) and antisymmetric (right) solutions for a Gallium Arsenide wire: the mean quadratic displacement 
    $\langle\hat{\xi}^{2}\rangle$ [panels (a) and b)], the mean quadratic velocity 
    $\langle\hat{\dot{\xi}}^{2}\rangle$ [panels (c) and (d)], and the energy current density $\langle\hat{J}\rangle$ [panels (e) and (f)]. The wire considered has thickness h = 200 nm, width W = 300 nm, and is 10 $\mu$m long. In each panel, five values of the temperature difference $\Delta T$ are considered: $10^{-5}$K (green), 0.1K (blue), 1K (red), and 10K (black) for $T_{1}=11$K, and $\Delta T = 10^{-5}$K (orange) for $T_{1}=10^{-4}$K. All panels share the same abscissa: the dimensionless position $y/W$ along the width of the wire.}
    \label{fig:meanValues}
\end{figure*}

In what follows, we solve numerically the just-described thermodynamic mean values that we refer to as observables for a Gallium Arsenide (GaAs) wire with a rectangular cross-section. The wire considered has thickness h = 200 nm, width W = 300 nm, and is 10 $\mu$m long. Because our theoretical approximations rely on small h/W ratios, we have verified that the observables’ transversal profile is more or less insensitive to increasing thickness. Only conductance is affected by becoming more sensitive to temperature, given the dependence of the dispersion relation on thickness. We consider five temperature configurations for the reservoirs. First, we set $T_{1}=11\mathrm{K}$ and varies $T_{2}$ to obtain $\Delta T= 10^{-5}$K, 0.1K, 1K, and 10K. For the last one we set $T_{1}=10^{-4}\mathrm{K}$ and $\Delta T= 10^{-5}$K. These are usual dimensions and conditions in mesoscopic thermal transport experiments~\cite{fon2002phonon,Tighe}. We consider the first cut-off frequency, $\omega_{0} = 0$, for the symmetric solution. For the antisymmetric solution, we consider the first non-zero cut-off frequency $\omega_{1}\sim1.12$~MHz, since the antisymmetric amplitude vanishes for $\omega_{0} = 0$. The number of modes is set to $N=30$ modes so that the frequency $\omega\left( k_{\ell}\right)$ does not overlap with the following cut-off frequency.

In Fig.~\ref{fig:meanValues}, we show the results for each observable's transversal profile, which extends along the wire ($x$-direction). The mean quadratic displacement in Eq.~\eqref{eq:meanQuadraticDisplacement} is shown in panels (a) and (b) of Fig.~\ref{fig:meanValues}; the mean quadratic velocity in Eq.~\eqref{eq:mqv} is shown in panels (c) and (d) of the same figure; finally, the profile of the energy current density, Eq.~\ref{eq:energyCurrentDensity}, is depicted in Figs.~\ref{fig:meanValues} (e) and \ref{fig:meanValues} (f). Notice that each observable is a definite positive quantity, and they are even functions of $y/W$, with respect to the origin at the middle of the wire along the transverse direction ($y/W=0$). Also, we notice that while the observables have a single valley in the middle for the symmetric solutions, they have three valleys for the antisymmetric solutions. This is because the later solutions have a nonzero cut-off frequency and, therefore, have more energy; this is an analogy to a string; the vibration mode has more nodes as the frequency increases. 

\begin{figure}
    \centering
    \includegraphics[width=1.0\columnwidth]{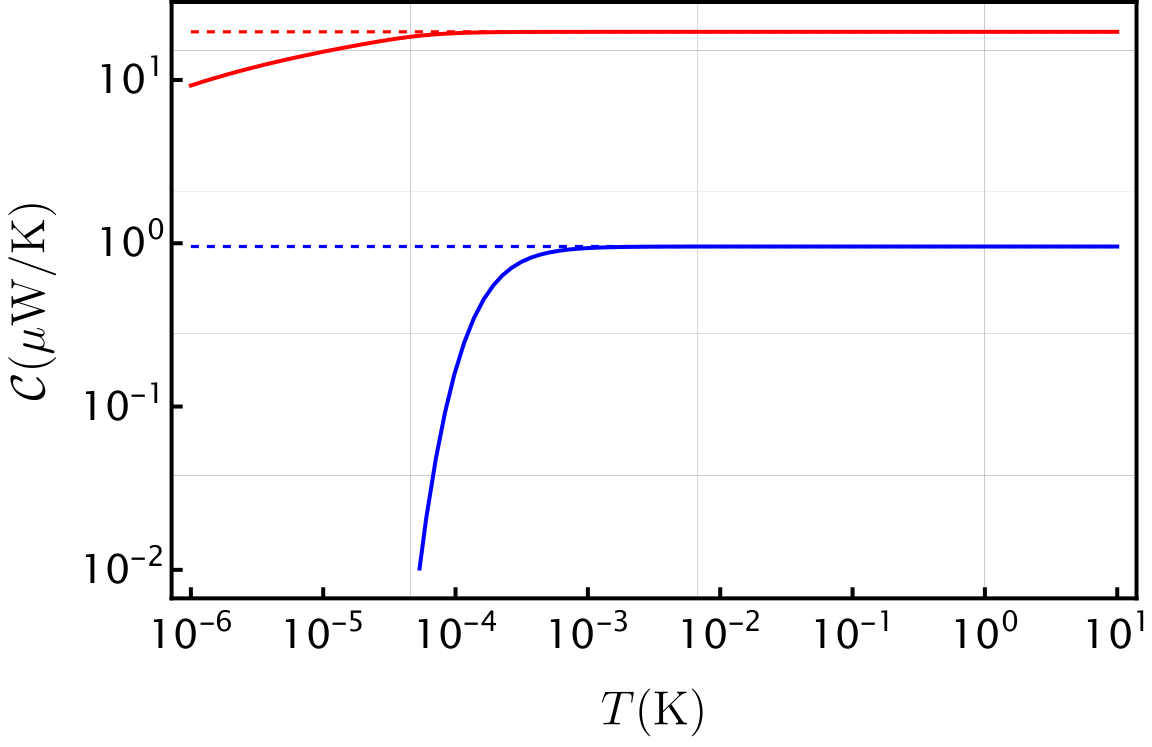}
    \caption{Log-log plot of the thermal conductance $\mathcal{C}$ as a function of the temperature for the symmetric (red) and antisymmetric (blue) modes. For $T>1$mK, approximately, the thermal conductance remains constant. The red dashed line is $\mathcal{C} = 19.7 \mu\mathrm{W/K}$; the blue dashed line corresponds to $\mathcal{C} = 0.955\mu\mathrm{W/K}$.}
    \label{fig:con}
\end{figure}

On the other side, the free-standing boundary conditions are responsible for the highest values of the observables at the edge of the wire in both kinds of solutions. Interestingly, the magnitude of an observable in the symmetric case is larger than the corresponding one in the antisymmetric case. That is, the symmetric solutions contribute mainly to the energy transport, as can be observed in Fig.~\ref{fig:con}, where the thermal conductance is plotted as a function of temperature. Not surprisingly, we also note from Fig.~\ref{fig:meanValues} that the magnitude of the energy current density increases as the temperature difference increases. The other observables follow behavior in the inverse sense. This change is because the displacement and velocity mean values depend on the sum of the particle populations for different energy levels. In contrast, the energy current density depends only on the difference of the populations.  

The thermal conductance is shown in Fig.~\eqref{fig:con} as a function of the temperature. It has a similar behavior for both kinds of solutions, but as we just mentioned above, there is a higher contribution to thermal transport from the symmetric modes. Also, we note that the thermal conductance tends quickly to its asymptotic value at high temperatures, indicating phonon saturation. In the regime of small temperature values, the thermal conductance tends to zero very fast, much faster in the antisymmetric case. These two behaviors with respect to temperature dependence are pretty similar to that of the intrinsic conductance of a one-dimensional mesoscopic resistor~\cite{Gao}. 

\section{Conclusions}
\label{sec:4}

Using the second quantization framework, we studied the phonon transport of a two-dimensional elastic waveguide's symmetric and antisymmetric modes between two thermal reservoirs. To do that, we solved the classical Kirchhoff-Love equation for rectangular wires and explored the dispersion relation for both kinds of solutions. We obtained the mean quadratic displacement transversal profile, which allows us to calculate the energy density current and the temperature field. Also, we have calculated the conductance for a small temperature difference. We found that the profiles of these observables are different for the symmetric and antisymmetric modes, but the symmetric modes contribute more than the antisymmetric ones to the thermal transport. Thermal conductance has an ohmic behavior, like a one-dimensional resistor, but for any temperature. Moreover, the behavior of the thermal conductance with respect to the temperature reaches a phonon saturation at high temperatures and decays rapidly as the temperature decreases. This result is quite similar to what happens to the electrical conductance of one-dimensional mesoscopic conductors. We expect that our findings help in the understanding of phonon-mediated thermal transport in two-dimensional mesoscopic quantum devices and that our approach can be successfully extended to describe thermal transport through disordered wires, too.

\begin{acknowledgments}
GR is financially supported by the PhD scholarship of CONAHCyT, under Contract No. 777351. MABM acknowledges financial support from the PEAPDI 2024 project from the DCBI UAM-I. EBR acknowledges ``Estancias posdoctorales por M\'exico 2023" program from CONAHCYT M\'exico for postdoctoral scholarship and acknowledges MSc. Adair Campos-Uscanga for numerical methods discussions.
\end{acknowledgments}

\section*{Author Declarations}

\subsection*{Conflict of Interest}

The authors have no conflicts to disclose.

\subsection*{Author Contributions}

All authors contributed equally to the manuscript. All authors have read and 
agreed to the published version of the manuscript.

\section*{Data Availability Statement}

The data that support the findings of this study are available from the 
corresponding author upon reasonable request.

\bibliography{bibliography}

\end{document}